\documentstyle[preprint,eqsecnum,aps,graphics]{revtex}

\begin{document}

\draft

\title{Proton capture cross section of Sr isotopes and their importance 
for nucleosynthesis of proton-rich nuclides}
\author{Gy. Gy\"urky, E. Somorjai, Zs. F\"ul\"op}
\address{
Institute of Nuclear Research of the Hungarian Academy of
Sciences (ATOMKI) Bem t\'er 18/c H-4001, Debrecen, Hungary
}
\author{S. Harissopulos, P. Demetriou}
\address{
Institute of Nuclear Physics, NCSR ``Demokritos'', 153.10 Aghia
Paraskevi, Athens, Greece
}
\author{T. Rauscher}
\address{
Dept.\ f\"ur Physik und Astronomie, Univ.\ Basel,
Klingelbergstr.\ 82,
4056 Basel, Switzerland\\
Dept.\ of Astronomy and Astrophysics, UCSC, Santa Cruz, CA
95064, USA
}
\date{\today}
\maketitle
\begin{abstract}
The (p,$\gamma$) cross sections of three stable Sr isotopes have
been measured in the astrophysically relevant energy range. These
reactions are important for the $p$-process in stellar
nucleosynthesis and, in addition, the reaction cross sections in
the mass region up to 100 are also of importance concerning the
$rp$-process associated with explosive hydrogen and helium
burning. It is speculated that this $rp$-process could be
responsible for a certain amount of $p$-nuclei in this mass
region.  The (p,$\gamma$) cross sections of $^{84,86,87}$Sr
isotopes were determined using an activation technique. The
measurements were carried out at the 5~MV Van de Graaff
accelerator of the ATOMKI, Debrecen. The resulting cross sections
are compared with the predictions of statistical model calculations.
The predictions are in good agreement with the experimental results
for $^{84}$Sr(p,$\gamma$)$^{85}$Y  whereas the other 
two reactions exhibit differences
that increase with mass number. The corresponding astrophysical
reaction rates have also been computed.
\end{abstract}


\section{Introduction}

Most of the stable proton rich nuclides with charge number Z $\geq$ 34 are
the so-called $p$-nuclei.  Their production by the $s$- or $r$-process
is blocked by stable nuclei and hence they are 10 to 100
times less abundant than the more neutron-rich isotopes. The main
stellar mechanism synthesizing these neutron-deficient nuclei is
a process called $\gamma$- or $p$-process \cite{woo78,ray95} involving a series
of ($\gamma$,n), ($\gamma$,p) and ($\gamma$,$\alpha$) reactions on pre-existing
more neutron-rich nuclei. The most favourable site for such a process to occur
is the O/Ne-layers of massive stars during presupernova phase 
\cite{ray95,how91,arn76}
or during their explosion as supernovae type II \cite{ray95,wooXX,rau95,arn92}.
Recent investigations of the $rp$-process associated with explosive hydrogen
and helium burning in X-ray bursters have shown that it could also
produce $p$-nuclei in the mass region up to 100. However, it is yet
unclear if and in what amount these nuclides could be ejected into the
interstellar medium \cite{wal81,schatz01}.

Describing the synthesis of p-nuclei and calculating their
abundances requires an extended  reaction network calculation
involving more than 10000 reactions, many of which are
photodisintegrations releasing protons, the inverse of radiative
proton capture (see e.g.\ \cite{ray90,heg00,rau_apj}).
In contrast to neutron capture reactions which are comparatively well
studied over the complete mass range of stable isotopes (see, e.g.\
\cite{Bao2000}), low-energy charged particle reaction data are
scarce for the mass region above iron.
Hence, so far, nucleosynthesis calculations of processes involving
charged particles (either by directly interacting with target nuclei
or as a result of photodisintegration) are based on
theoretical predictions of the Hauser-Feshbach model
(e.g.\ \cite{nonsmokerrates,gor2000,nonsmokerrates2}).
Therefore, a
systematic experimental study of charged particle reactions was
initiated in several laboratories in order to obtain an extensive and reliable
experimental data base of proton and $\alpha$-capture cross sections
\cite{ful96,sau97,som98,bork98,chlo99}. Considerable differences with the
theoretical predictions were found for certain ($\alpha$,$\gamma$) reactions
while the measured proton capture cross sections exhibited less differences.
The $\alpha$ optical model potential at low energies was identified as the
main source of the difference \cite{mohr1997} and this also led to the
idea of systematically studying the low-energy $\alpha$ optical 
model potential in (n,$\alpha$) reactions
\cite{gled2000,koe2001}, whereas the proton optical model potential
is considered to be more reliable \cite{nonsmokerlev,nonsmokerrates}. However,
due to the rareness of experimental data in the intermediate and heavy mass
range we are still far from obtaining complete 
systematics which would provide the basis
for a thorough test of the theoretical descriptions and therefore it is
necessary to pursue further investigations of (p,$\gamma$) reactions close
to the astrophysically relevant energy range.

In this work we present for the first time measurements of the
(p,$\gamma$) cross sections of three stable Sr isotopes, $^{84,
86, 87}$Sr, in the astrophysically relevant energy range using an
activation technique. Our results are compared with
statistical model (Hauser-Feshbach) calculations.

\section{Investigated reactions}

The element Sr has four stable isotopes with mass numbers A~=~84, 86, 87, and
88, having isotopic abundances of 0.56\%, 9.86\%, 7.00\%, and
82.58\%, respectively. Only the (p,$\gamma$) cross
section of the first 3 isotopes can be determined by activation because
in the case of $^{88}$Sr(p,$\gamma$) reaction the product nucleus 
$^{89}$Y is stable
and its isomer is short lived (T$_{1/2}$=16.06s). In the case
of $^{84}$Sr and $^{86}$Sr the partial cross sections leading to
the isomer and ground state of the corresponding Y isotopes can be
determined separately because of the different decay pattern of the
isomeric and ground state.

The relevant part of the chart of nuclides can be seen in
Fig.\ref{nuclide} where the decay of the reaction products can also be seen.
The decay parameters used for the analysis are summarized in Table
\ref{decay}.

\section{Experimental procedure}

\subsection{Target properties}

As mentioned above, the isotope $^{88}$Sr which cannot be
investigated by activation, has the highest natural abundance of
about 80\%. Thus, in case of natural targets only 20\% of the
material is effective for the activation. However, the use of
natural Sr targets has the advantage that the isotopic abundances
are very well known and natural
 Sr is easily available in many chemical forms.

In our measurements we used natural SrF$_2$ targets, evaporated
onto thick carbon backings. The fluorine content of the target causes
no disturbance because in the investigated energy range there is
no activity produced by proton bombardment on F and it has a low
mass number, thus it is well separated from Sr in Rutherford
Backscattering (RBS) spectra that were also measured during the 
irradiations in order to monitor the target stability as described below.
Carbon causes no
disturbing long-lived activity and due to its low mass number,
the C edge lies far below the Sr and F peaks in the RBS spectra.

The number of target atoms was determined by proton induced X-ray
emission (PIXE) at the PIXE set-up of the ATOMKI \cite{PIXE}. The
results were checked by $\alpha$-RBS determining the width of the
Sr peak. In most cases we found good agreement. The method in ref.
\cite{sau97} which compares the area of the Sr peak and the height
of the backing (carbon) edge in the proton-RBS spectrum is not
applicable because of the non-Rutherford behavior of p-C elastic
scattering. Altogether 33 targets were prepared, some of them were
only used for test runs.

Using natural Sr targets the (p,$\gamma$) cross section of
$^{84}$Sr, $^{86}$Sr and $^{87}$Sr can be determined
simultaneously in a single activation procedure. At E$_p$=2.67~MeV
bombarding energy the $^{87}$Sr(p,n)$^{87}$Y channel opens, which
results in the same product nucleus as $^{86}$Sr(p,$\gamma$).
Consequently, above this energy the cross section of
$^{86}$Sr(p,$\gamma$) reaction cannot be deduced. However, this is
the only disturbing proton induced reaction channel on Sr isotopes
which is open in the investigated energy range.

\subsection{Activation}

The activations were carried out at the 5~MV Van de Graaff
accelerator  of the ATOMKI. The energy range from E$_p$=1.5
to 3~MeV was covered with 100~keV steps.
The schematic view of the target
chamber can be seen in Fig.\ref{chamber}. After the last beam
defining aperture the whole chamber served as a Faraday-cup to
collect the accumulated charge. A secondary electron suppression
voltage of $-300$ V was applied at the entrance of the chamber. The
irradiations lasted between 6 and 24 hours with a beam current of
typically 5 to 10 $\mu$A. Thus, the collected charge varied between
120 and 650 mC. The current was kept as stable as possible but to
follow the changes the integrator counts were recorded in
multichannel scaling mode, stepping the channel in every minute.

A surface barrier detector was built into the chamber at
$\Theta$=150$^\circ$ relative to the beam direction to detect the
backscattered protons and this way to monitor the target stability.
The RBS spectra were taken continuously and stored regularly
during the irradiation. In those cases when target deterioration
was found, the irradiation was repeated with another target. Fig.
\ref{rbs} shows a typical RBS spectrum.

The beam was wobbled across the last diaphragm to have a uniformly
irradiated spot of diameter of 8~mm on the target. The target
backing was directly water cooled with an isolated water
circulating system.

Between the irradiation and $\gamma$-counting, a waiting time of 1
to 2 hours was inserted in order to let the disturbing short lived
activities decay out.

\subsection{Detection of induced $\gamma$-radiation}

The $\gamma$-radiation following the $\beta$-decay of the produced
Y  isotopes was measured with a HPGe detector of 40\% relative
efficiency. The low intensity gamma radiation necessitated the use
of a very close geometry: the target was mounted in a holder
directly onto the end of the detector cap. The whole system was
shielded by 10~cm thick lead against laboratory background.

The $\gamma$-spectra were taken for at least 48 hours and stored
regularly  in order to follow the decay of the different reaction
products.

The efficiency curve of the detector was determined with
$^{133}$Ba, $^{56}$Co and $^{152}$Eu sources in the same close
geometry. The measured points were fitted with a third degree
logarithmic polynomial to yield the efficiency curve for the whole
energy region of interest. The efficiency measurements were
checked with Monte Carlo simulations and good agreement was found
in the whole energy range \cite{mohr}.

Fig. \ref{gammaspec} shows an off-line $\gamma$-spectrum taken
after irradiation with 3~MeV protons. $\gamma$-lines used for the
analysis are indicated by arrows.

\section{Experimental results}

Tables \ref{84tab},\ref{86tab},\ref{87tab} summarize the
experimental cross sections of the three Sr isotopes, while Figs.
\ref{84fig},\ref{86fig},\ref{87fig} show the derived astrophysical
$S$--factors as a function of center-of-mass energy. In the figures,
the predictions of the Hauser-Feshbach statistical model codes
MOST \cite{most} and NON-SMOKER \cite{nonsmoker,nonsmokerrates2} can
also be seen.

\subsection{$^{84}$Sr(p,$\gamma$)$^{85}$Y$^{g,m}$}

Proton capture of $^{84}$Sr populates the ground
(T$_{1/2}$~=~2.68~h) and isomeric (T$_{1/2}$~=~4.86~h) state of
$^{85}$Y. The isomeric state decays directly to $^{85}$Sr without
internal transition to the ground state. In principle, the cross
sections to the isomer (m) and ground (g) state can be determined
independently using the 231.65~keV and 504.4~keV lines. However,
at low bombarding energies the intensity of the latter transition
is very low compared to the strong 511~keV annihilation line.
Therefore the subsequent decay of $^{85}$Sr had to be investigated. This
decay results in the emission of 151.16~keV radiation with high
intensity only when the ground state of the original $^{85}$Y was
populated. Thus by detecting this line, the ground  and the isomeric
state can be distinguished.

Here we should point out a discrepancy observed among the decay
data in the literature \cite{sie91} (see Table \ref{decay}). At
the highest energies the above analysis can be performed
independently with both methods: with detection of the direct
231.65~keV and 504.4~keV lines and with the delayed 151.16~keV
transition. The results from the two methods were found to be
different by a factor of about 2. This indicates that at least one
of the adopted relative $\gamma$-ray intensity values is
incorrect. A test measurement was performed in order to resolve
this discrepancy. The precision of this measurement was not enough
to provide new values to these $\gamma$-ray intensities, however,
we could determine which data are incorrect. We found that the
relative intensity of the 151.16~keV line is wrong. Thus the
analysis was performed by using the 231.65~keV and 504.4~keV lines
and the 151.16~keV line was used only for relative measurements
where the weak 504.4~keV line could not be detected with sufficient
precision.

Fig. \ref{84fig} shows the measured and calculated $S$--factor
values of the $^{84}$Sr(p,$\gamma$)$^{85}$Y reaction. The results
of both model calculations are in satisfactory agreement with the
measured values.

\subsection{$^{86}$Sr(p,$\gamma$)$^{87}$Y$^{g,m}$}

For this isotope the two partial cross sections can be derived
easily since the 380.79~keV $\gamma$-line comes entirely from the
internal transition of the isomer.

As mentioned above, at E$_p$=2.67~MeV bombarding energy the
$^{87}$Sr(p,n) channel opens, hence above this energy only the sum
of the cross section of $^{86}$Sr(p,$\gamma$)$^{87}$Y and
$^{87}$Sr(p,n)$^{87}$Y reactions (weighted with the isotopic
abundances) can be deduced. However, calculations show that in the
investigated energy region the (p,n) channel is much (about two
orders of magnitude) weaker than the (p,$\gamma$). Hence the
points above the threshold are included in Tab. \ref{87tab}
and Fig. \ref{86fig} although in brackets.

The resulting cross sections are significantly lower than the
values of the model calculations.

\subsection{$^{87}$Sr(p,$\gamma$)$^{88}$Y}

The isomer of $^{88}$Y is very short-lived (T$_{1/2}$~=~13.9~ms)
and it decays with internal transitions to the ground state. Thus,
with the activation technique the total proton capture cross sections
can be determined. The analysis was carried out with the two
cascading $\gamma$-rays from the $^{88}$Y decay.

The statistical model calculations strongly overestimate the
measured data as can be seen in fig. \ref{87fig}.

\section{Discussion}

\subsection{Dependence on nuclear properties}

The comparison of the $S$--factor data with the theoretical predictions seems
to exhibit a trend depending on the nuclear mass: the more heavy and
neutron-rich the measured Sr isotopes become, the more the cross
sections are overestimated by the statistical model calculations. 
The differences between the theoretical models is less grave, they differ by
about 30\%, nearly independently of mass and projectile energy. While
most of the descriptions of nuclear properties used in the two codes
are similar, the nuclear level densities are different. The
standard NON-SMOKER calculations use a global level density based on the 
shifted Fermi-gas model \cite{nonsmokerlev}, while 
the MOST code makes use of a microscopic global model level
density based on a Hartree-Fock-BCS method \cite{gor96,dem01}. 
This may explain the difference in the two
results. It is evident that both predictions seem to become worse for isotopes
approaching the closed neutron shell $N=50$. The level densities also depend
on microscopic properties like pairing and shell effects. It is known
\cite{nonsmokerlev,rau98,pea96,moe95} that those are often predicted 
inaccurately at closed shells, often overestimating the effect.

Apart from the nuclear level densities, the proton optical model 
potential also plays a crucial role.
Further studies of the optical potential dependence in comparison to the
measured data are shown in Figs.\ \ref{86comp}, \ref{87comp} for the reactions
$^{86}$Sr(p,$\gamma$)$^{87}$Y and 
$^{87}$Sr(p,$\gamma$)$^{88}$Y. In both figures
we plot the $S$--factor data and the values obtained with the standard
NON-SMOKER and MOST potential JLM \cite{jlm}, 
an equivalent square well potential (ESW)
of \cite{HWFZ}, and the two Saxon-Woods type phenomenological potentials 
of PER \cite{perey} and BEC \cite{becch}. While the ESW
potential and the PER potential do not reproduce the energy dependence of
the $S$--factors well, it is reproduced by the BEC potential more
accurately. Among the 4 potentials considered, the JLM potential leads
to the best overall description of the energy dependence of the 
experimental data, except for the 
low-energy region where the ESW potential does 
better. The above findings hold for all the reactions considered in this work.

\subsection{Astrophysical reaction rates}

It is premature to try to derive an improved global proton potential
from these three measured isotopes and the scarce data available for
other elements. From the results of Sec. V. A. 
it is clear that an optical potential can be found to reproduce 
the data presented in this work. Unfortunately, such a 
potential would only describe these reaction data and will not be a global one.
In order to calculate the integral for the astrophysical reaction rate with 
sufficient numerical accuracy it is only necessary to reproduce 
the experimental data by the theoretical calculation well. 
This can be achieved by dividing the results 
obtained with the JLM potential and shown in Figs. \ref{86comp}, \ref{87comp} 
by a constant
factor of 1.7 for the reaction $^{86}$Sr(p,$\gamma$)$^{87}$Y and a
factor of 2.2 for the reaction $^{87}$Sr(p,$\gamma$)$^{88}$Y,
respectively. The astrophysical reaction rates obtained from those
renormalizations are given in Tab.\ \ref{ratetable}. A similar renormalization 
procedure was used for the reaction 
$^{84}$Sr(p,$\gamma$)$^{85}$Y. The fit to the experimental data quoted in 
Tab.\ \ref{ratetable} produced values which lie halfway between the 
NON-SMOKER and MOST results, yielding mean cross sections and rates which are 
equivalent to a renormalization of the MOST values by a factor of 1.15 and a 
downscaling of the NON-SMOKER values by the appropriate factor. 
The extensions to lower and higher 
temperatures are obtained by renormalizing the NON-SMOKER values 
\cite{nonsmokerrates2} by the above values.

In addition to the laboratory rates obtained 
when the target is in the ground state,
the stellar rates calculated for a thermally excited target are also
shown in Tab.\ \ref{ratetable}. As the measured reactions only involve
targets in the ground state, it was assumed that the stellar rates
scale like the laboratory rates.
By re-adjustment of the reaction rate parametrization of
Ref.\ \cite{nonsmokerrates} to the new data, we arrive at new values for the
parameters $a_0$ of the stellar rate:
250.84 for $^{84}$Sr(p,$\gamma$)$^{85}$Y, 245.13 for
$^{86}$Sr(p,$\gamma$)$^{87}$Y, and 208.87 for
$^{87}$Sr(p,$\gamma$)$^{88}$Y. All other fit parameters remain the same
since the energy dependence is reproduced well by the theory.

In general, it should be noted that even modern {\em global} statistical model
predictions -- which are not locally tuned to experimentally known
nuclear properties -- bear an uncertainty of a factor of 1.3--2.0. Thus,
the deviations from experimental data found in this work are not surprising and
still within the expected uncertainties. 
Nevertheless, the observed discrepancies
underline the importance of carrying out experimental studies in order to test
the reliability of Hauser-Feshbach and improve the accuracy of the calculated
reaction rates used in astrophysical applications.

\section{Summary}

We have measured proton capture cross sections on the Sr isotopes with
mass numbers 84, 86, and 87 in the astrophysically important energy
range of $1.5\leq E_{\rm c.m.}\leq 3$ MeV (this corresponds to a
temperature range of about $T\simeq (1.0-4.0)\times 10^9$ K). While we
find good agreement between experiment and theory for the reaction
$^{84}$Sr(p,$\gamma$)$^{84}$Y, the predictions for the other two
reactions differ considerably from our results although not as much
as in the case of previous $\alpha$
capture measurements. The reason for these inaccuracies could be attributed
to uncertainties in the optical potentials and nuclear level densities used
in the statistical model calculations. Although the uncertainties in
global reaction rate calculations are expected to be of the magnitude
found in the current investigation, further investigation is required in order
to resolve the discrepancies and arrive at improved 
global predictions of nuclear properties. The newly derived reaction rate can 
be directly used in astrophysical applications.

\acknowledgements

This work was supported by the Hungarian National Scientific Fund
(OTKA T034259) and the Hungarian-Greek collaboration (T\'eT GR 23/99). T. R.
is a PROFIL professor (Swiss National Science Foundation grant 2124-055832.98).
This work was partially supported by the Swiss NSF (grant 2000-061822.00) and 
the U.S. NSF (contract NSF-AST-97-31569). Zs. F. is a Bolyai fellow.

\begin{table}
\caption{Decay parameters of the Y product nuclei taken from the
literature}
\begin{tabular}{ccccc}
\hline
\parbox[t]{2.2cm}{\centering{Product \\ nucleus}} &
\parbox[t]{2.2cm}{\centering{Half life [hour]}} &
\parbox[t]{2.2cm}{\centering{Gamma \\energy [keV]}} &
\parbox[t]{2.2cm}{\centering{Relative \\ intensity \\ per decay [\%]}} &
\parbox[t]{2.2cm}{\centering{Reference}} \\
\hline $^{85}$Y$^g$ & 2.68 $\pm$ 0.05 & 231.65 & 84 $\pm$ 8 &
\cite{sie91}
\\
             &                 & 504.4  & 60 $\pm$ 4     &    \\
             &                 & 151.16  & 12.9 $\pm$ 0.3     &    \\
&&&&\\ $^{85}$Y$^m$ & 4.86 $\pm$ 0.13 & 231.67 & 22.8 $\pm$ 2.2 &
\cite{sie91} \\
             &                 & 504.4  & 1.5 $\pm$ 0.1     &    \\
&&&&\\ $^{87}$Y$^g$ & 79.8 $\pm$ 0.03 & 388.53 & 82.1 $\pm$ 0.5 &
\cite{sie91a} \\
             &                 & 484.81 & 89.7 $\pm$ 0.6 &    \\
&&&&\\ $^{87}$Y$^m$ & 13.37 $\pm$ 0.03 & 380.79 & 78.1 $\pm$ 0.1 &
\cite{sie91a} \\ &&&&\\ $^{88}$Y$^g$ & 106.7 $\pm$ 0.04 & 898.04 &
98.6 $\pm$ 0.3 & \cite{mul88} \\
             &       day          &1836.06 & 99.2 $\pm$ 0.3 &    \\
\hline
\end{tabular}
\label{decay}
\end{table}

\begin{table}
\caption{Experimental cross section of the
$^{84}$Sr(p,$\gamma$)$^{85}$Y reaction}
\begin{tabular}{cccc}
E$_{\rm c.m., eff.}$ & \multicolumn{3}{c}{Cross section [$\mu$barn]}
\\ \cline{2-4}  \, [keV] & Ground state & Isomer & Total \\ \hline
2962 & 574.8 $\pm$ 47.7 & 424.6 $\pm$ 99.7 & 999.4 $\pm$ 147.4 \\
2836 & 210.5 $\pm$ 18.7 & 372.5 $\pm$ 65.8 & 583.0 $\pm$ 84.5 \\
2764 & 133.5 $\pm$ 11.9 & 251.9 $\pm$ 47.5 & 385.4 $\pm$ 59.4 \\
2664 & 140.0 $\pm$ 12.8 & 228.1 $\pm$ 44.1 & 368.1 $\pm$ 56.9 \\
2566 & 95.4 $\pm$ 8.4 & 161.9 $\pm$ 30.3 & 257.3 $\pm$ 38.7 \\
2468 & 87.0 $\pm$ 7.3 & 119.9 $\pm$ 28.0 & 206.9 $\pm$ 35.3 \\
2368 & 67.6 $\pm$ 6.0 & 101.6 $\pm$ 20.2 & 169.2 $\pm$ 26.2 \\
2267 & 34.5 $\pm$ 2.9 & 47.8 $\pm$ 11.6 & 82.3 $\pm$ 14.5 \\ 2170
& 29.6 $\pm$ 2.5 & 45.6 $\pm$ 10.7 & 75.2 $\pm$ 13.2 \\ 2071 &
18.5 $\pm$ 1.7 & 19.5 $\pm$ 5.2 & 38.0 $\pm$ 6.9 \\ 1973 & 13.0
$\pm$ 1.2 & 22.0 $\pm$ 5.3 & 35.0 $\pm$ 6.5 \\ 1871 & 7.54 $\pm$
0.7 & 10.9 $\pm$ 2.7 & 18.4 $\pm$ 3.4 \\ 1774 & 4.39 $\pm$ 0.4 &
5.22 $\pm$ 1.5 & 9.61 $\pm$ 1.9 \\ 1673 & 0.851 $\pm$ 0.27 & 2.30
$\pm$ 0.9 & 3.15 $\pm$ 1.2\\
\end{tabular} \label{84tab}
\end{table}

\begin{table}
\caption{Experimental cross section of the
$^{86}$Sr(p,$\gamma$)$^{87}$Y reaction}
\begin{tabular}{cccc}
E$_{\rm c.m., eff.}$ & \multicolumn{3}{c}{Cross section [$\mu$barn]}
\\ \cline{2-4}  \, [keV] & Ground state & Isomer & Total \\ \hline
2963 & (496.0 $\pm$ 39.7) & (136.5 $\pm$ 10.2) & (632.5 $\pm$
49.9) \\ 2864 & (311.6 $\pm$ 27.0) & (98.0 $\pm$ 7.98) & (409.6
$\pm$ 35.0) \\ 2765 & (227.6 $\pm$ 19.4) & (57.9 $\pm$ 4.71) &
(285.5 $\pm$ 24.1) \\ 2665 & (172.4 $\pm$ 15.0) & (46.1 $\pm$
3.86) & (218.5 $\pm$ 18.9) \\ 2567 & 152.8 $\pm$ 12.8 & 38.8 $\pm$
3.13 & 191.6 $\pm$ 15.9 \\ 2468 & 128.4 $\pm$ 10.5 & 40.7 $\pm$
3.10 & 169.1 $\pm$ 13.6 \\ 2368 & 99.9 $\pm$ 8.52 & 31.7 $\pm$
2.58 & 131.6 $\pm$ 11.1 \\ 2268 & 52.0 $\pm$ 4.15 & 10.7 $\pm$
0.80 & 62.7 $\pm$ 4.95 \\ 2171 & 33.0 $\pm$ 2.65 & 10.3 $\pm$ 0.77
& 43.3 $\pm$ 3.42 \\ 2072 & 26.6 $\pm$ 2.06 & 8.39 $\pm$ 0.62 &
35.0 $\pm$ 2.68 \\ 1973 & 18.4 $\pm$ 1.43 & 5.11 $\pm$ 0.38 & 23.5
$\pm$ 1.81 \\ 1871 & 9.14 $\pm$ 0.72 & 2.13 $\pm$ 0.16 & 11.3
$\pm$ 0.88 \\ 1774 & 7.46 $\pm$ 0.59 & 1.49 $\pm$ 0.11 & 8.94
$\pm$ 0.70 \\ 1673 & 3.25 $\pm$ 0.25 & 0.462 $\pm$ 0.035 & 3.71
$\pm$ 0.29 \\ 1577 & 1.30 $\pm$ 0.10 & 0.206 $\pm$ 0.016 & 1.51
$\pm$ 0.12 \\ 1477 & 0.637 $\pm$ 0.05 & 0.095 $\pm$ 0.007 & 0.732
$\pm$ 0.06 \\
\end{tabular} \label{86tab}
\end{table}

\begin{table}
\caption{Experimental cross section of the
$^{87}$Sr(p,$\gamma$)$^{88}$Y reaction}
\begin{tabular}{cc}
E$_{\rm c.m., eff.}$ & Cross section \\ \,[keV] & [$\mu$barn]
\\ \hline 2963 & 629.2 $\pm$ 47.1 \\ 2864 & 340.3 $\pm$ 27.7 \\
2765 & 332.5 $\pm$ 27.1
\\ 2665 & 262.7 $\pm$ 22.0 \\ 2567 & 232.0 $\pm$ 18.7 \\ 2469 &
156.0 $\pm$ 12.0 \\ 2369 & 121.0 $\pm$ 9.90 \\ 2268 & 67.7 $\pm$
5.12 \\ 2171 & 48.6 $\pm$ 3.71 \\ 2072 & 30.1 $\pm$ 2.28 \\ 1973 &
15.7 $\pm$ 1.31 \\ 1871 & 8.18 $\pm$ 0.67 \\ 1775 & 5.04 $\pm$
0.42 \\ 1674 & 2.03 $\pm$ 0.19 \\ 1577 & 1.39 $\pm$ 0.16 \\
\end{tabular} \label{87tab}
\end{table}

\begin{table}
\caption{Astrophysical reaction rates in cm$^3\,$s$^{-1}\,$mol$^{-1}$
derived from the new data.}
\begin{tabular}{rrrrrrr}
&\multicolumn{2}{c}{$^{84}$Sr(p,$\gamma$)$^{85}$Y} &
\multicolumn{2}{c}{$^{86}$Sr(p,$\gamma$)$^{87}$Y} &
\multicolumn{2}{c}{$^{87}$Sr(p,$\gamma$)$^{88}$Y} \\
\multicolumn{1}{c}{T [10$^9$ K]}&
\multicolumn{1}{c}{lab.}&\multicolumn{1}{c}{star}&
\multicolumn{1}{c}{lab.}&\multicolumn{1}{c}{star}&
\multicolumn{1}{c}{lab.}&\multicolumn{1}{c}{star} \\
\hline
 0.10&7.40E$-$27&7.40E$-$27&4.92E$-$27&4.92E$-$27&3.80E$-$27&3.80E$-$27\\
 0.15&2.85E$-$21&2.85E$-$21&1.89E$-$21&1.89E$-$21&1.46E$-$21&1.46E$-$21\\
 0.20&8.85E$-$18&8.85E$-$18&5.88E$-$18&5.88E$-$18&4.55E$-$18&4.55E$-$18\\
 0.30&2.09E$-$13&2.09E$-$13&1.39E$-$13&1.39E$-$13&1.07E$-$13&1.07E$-$13\\
 0.40&1.16E$-$10&1.16E$-$10&7.71E$-$11&7.71E$-$11&5.95E$-$11&5.95E$-$11\\
 0.50&9.82E$-$09&9.82E$-$09&6.53E$-$09&6.53E$-$09&5.05E$-$09&5.05E$-$09\\
 0.60&2.97E$-$07&2.97E$-$07&1.98E$-$07&1.98E$-$07&1.53E$-$07&1.53E$-$07\\
 0.70&4.67E$-$06&4.67E$-$06&3.11E$-$06&3.11E$-$06&2.40E$-$06&2.40E$-$06\\
 0.80&4.51E$-$05&4.51E$-$05&3.00E$-$05&3.00E$-$05&2.32E$-$05&2.32E$-$05\\
 0.90&3.00E$-$04&3.00E$-$04&1.99E$-$04&1.99E$-$04&1.54E$-$04&1.54E$-$04\\
 1.00&1.50E$-$03&1.50E$-$03&9.94E$-$04&9.94E$-$04&7.68E$-$04&7.68E$-$04\\
 1.50&3.54E$-$01&3.54E$-$01&2.35E$-$01&2.35E$-$01&1.82E$-$01&1.82E$-$01\\
 2.00&9.03E+00&9.12E+00&6.00E+00&6.06E+00&4.64E+00&4.68E+00\\
 2.50&8.04E+01&8.26E+01&5.34E+01&5.49E+01&4.13E+01&4.24E+01\\
 3.00&4.00E+02&4.21E+02&2.66E+02&2.80E+02&2.05E+02&2.16E+02\\
 3.50&1.38E+03&1.50E+03&9.18E+02&9.94E+02&7.09E+02&7.68E+02\\
 4.00&3.74E+03&4.15E+03&2.49E+03&2.76E+03&1.92E+03&2.13E+03\\
 4.50&8.53E+03&9.56E+03&5.67E+03&6.35E+03&4.38E+03&4.91E+03\\
 5.00&1.71E+04&1.90E+04&1.14E+04&1.26E+04&8.77E+03&9.77E+03\\
 6.00&5.19E+04&5.27E+04&3.45E+04&3.50E+04&2.67E+04&2.70E+04\\
 7.00&1.22E+05&9.56E+04&8.12E+04&6.35E+04&6.27E+04&4.91E+04\\
 8.00&2.41E+05&1.19E+05&1.60E+05&7.94E+04&1.24E+05&6.14E+04\\
 9.00&4.19E+05&1.13E+05&2.78E+05&7.53E+04&2.15E+05&5.82E+04\\
10.00&6.62E+05&9.03E+04&4.40E+05&6.00E+04&3.40E+05&4.64E+04
\end{tabular}
\label{ratetable}
\end{table}

\begin{figure}
\caption{Part of the chart of nuclides showing the investigated
reactions and the decay scheme of reaction products. Stable
nuclides are indicated by bold squares.}
\label{nuclide}
\end{figure}

\begin{figure}
\caption{Schematic view of the target chamber and the data acquisition.}
\label{chamber}
\end{figure}

\begin{figure}
\caption{A typical RBS spectrum taken at E$_p$=2.0~MeV. The Sr and
F peaks and the carbon edge are indicated.} \label{rbs}
\end{figure}

\begin{figure}
\caption{Off-line $\gamma$-spectrum taken after irradiation with
3~MeV protons. The $\gamma$-lines used for the analysis are indicated
by arrows. All the other peaks correspond to either laboratory
background and activity induced on target impurities or lines
of Y isotopes which were not used for the analysis.}
\label{gammaspec}
\end{figure}

\begin{figure}
\caption{Measured and calculated astrophysical $S$--factor of the
$^{84}$Sr(p,$\gamma$)$^{85}$Y reaction} \label{84fig}
\end{figure}

\begin{figure}
\caption{Measured and calculated astrophysical $S$--factor of the
$^{86}$Sr(p,$\gamma$)$^{87}$Y reaction. The measured points above
the $^{87}$Sr(p,n)$^{87}$Y threshold are put in parentheses.}
\label{86fig}
\end{figure}

\begin{figure}
\caption{Measured and calculated astrophysical $S$--factor of the
$^{87}$Sr(p,$\gamma$)$^{88}$Y reaction.} \label{87fig}
\end{figure}

\begin{figure}
\caption{Potential comparison for $^{86}$Sr(p,$\gamma$)$^{87}$Y. Shown
are the measured $S$--factors (EXP) and calculations using different
potentials but otherwise unchanged nuclear input: PER \protect\cite{perey},
BEC \protect\cite{becch}, JLM \protect\cite{jlm}, and ESW
\protect\cite{HWFZ}.}
\label{86comp}
\end{figure}

\begin{figure}
\caption{Potential comparison for $^{87}$Sr(p,$\gamma$)$^{88}$Y. Shown
are the measured $S$--factors (EXP) and calculations using different
potentials but otherwise unchanged nuclear input: PER \protect\cite{perey},
BEC \protect\cite{becch}, JLM \protect\cite{jlm}, and ESW
\protect\cite{HWFZ}.}
\label{87comp}
\end{figure}

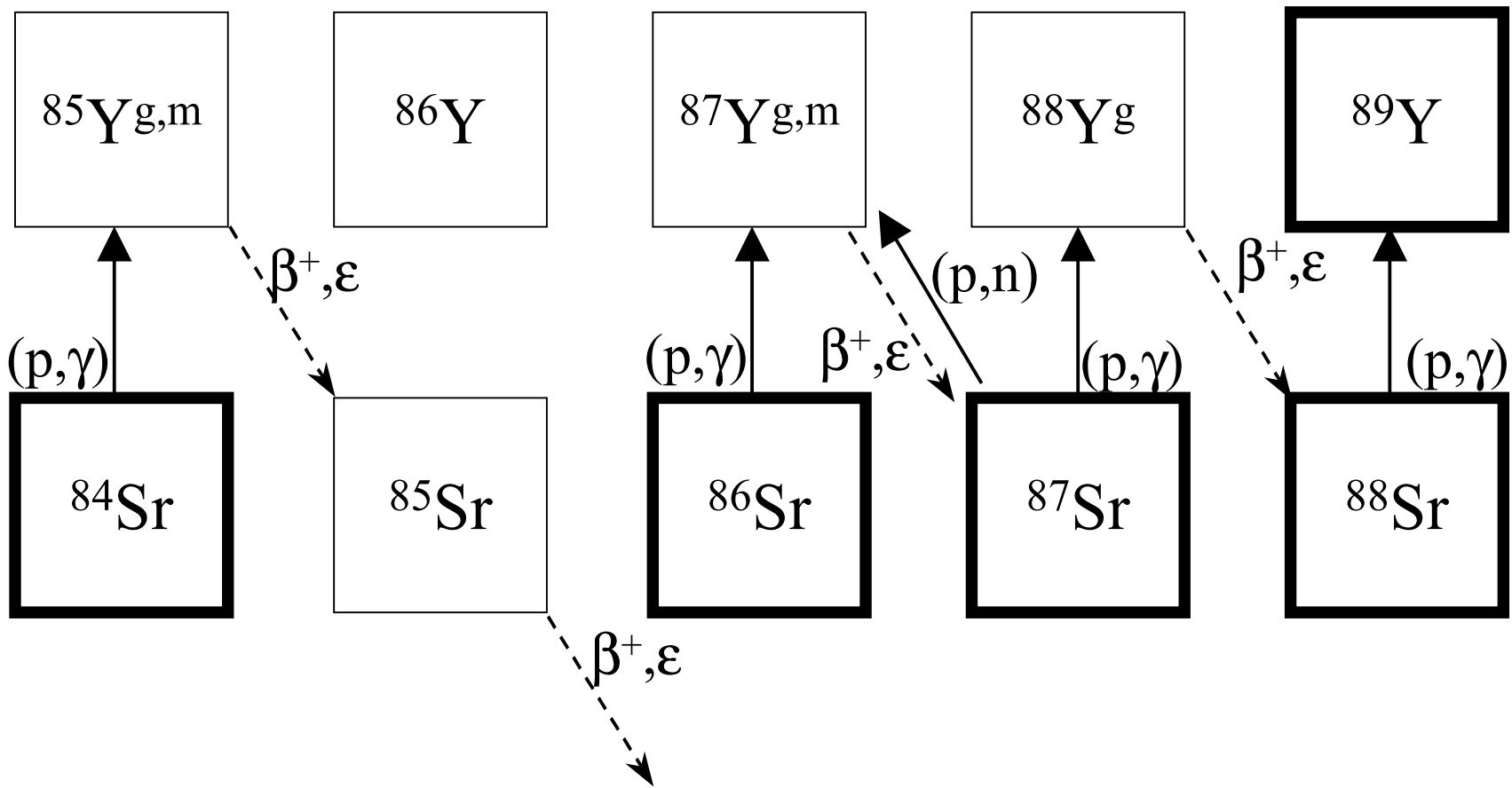
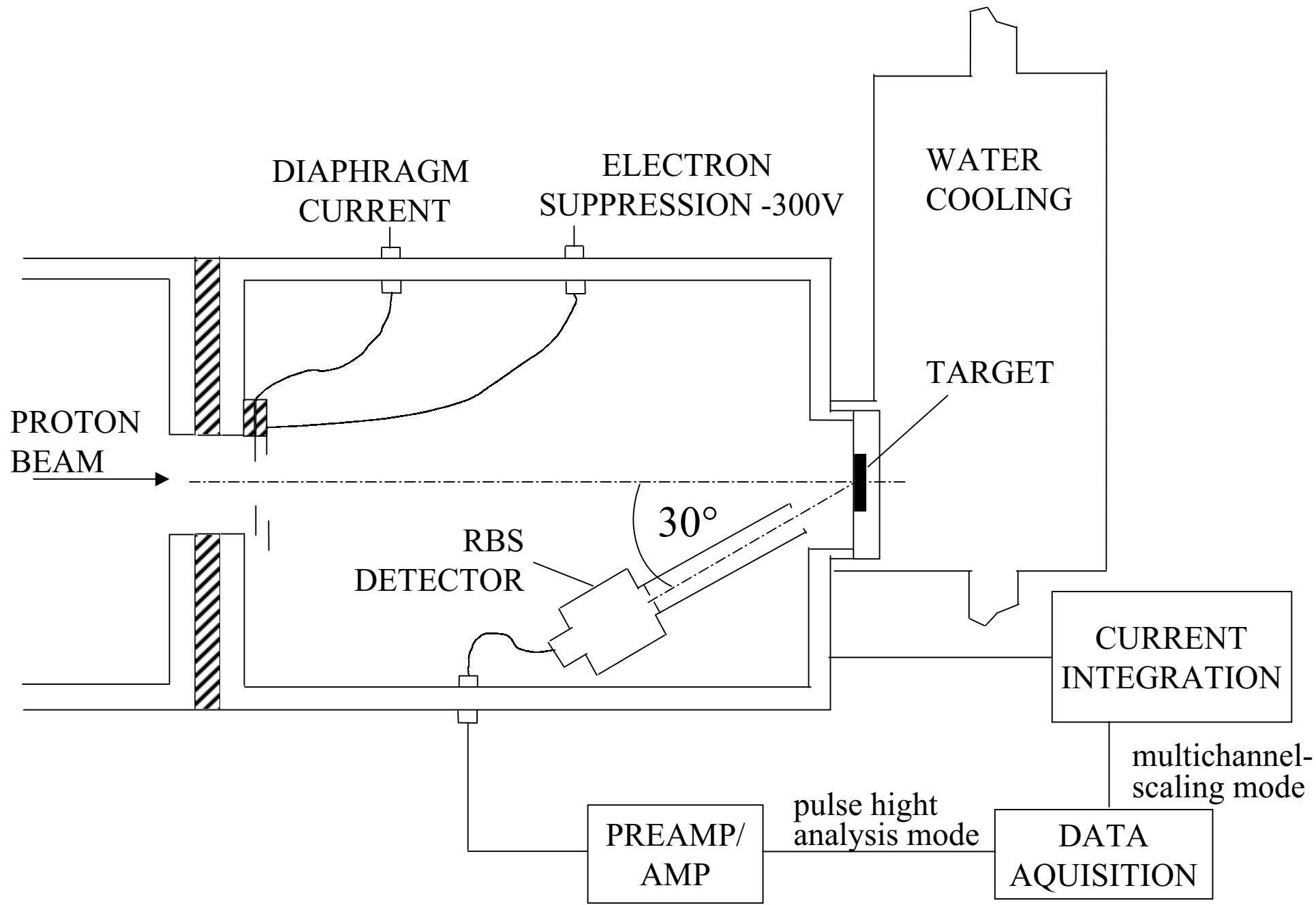
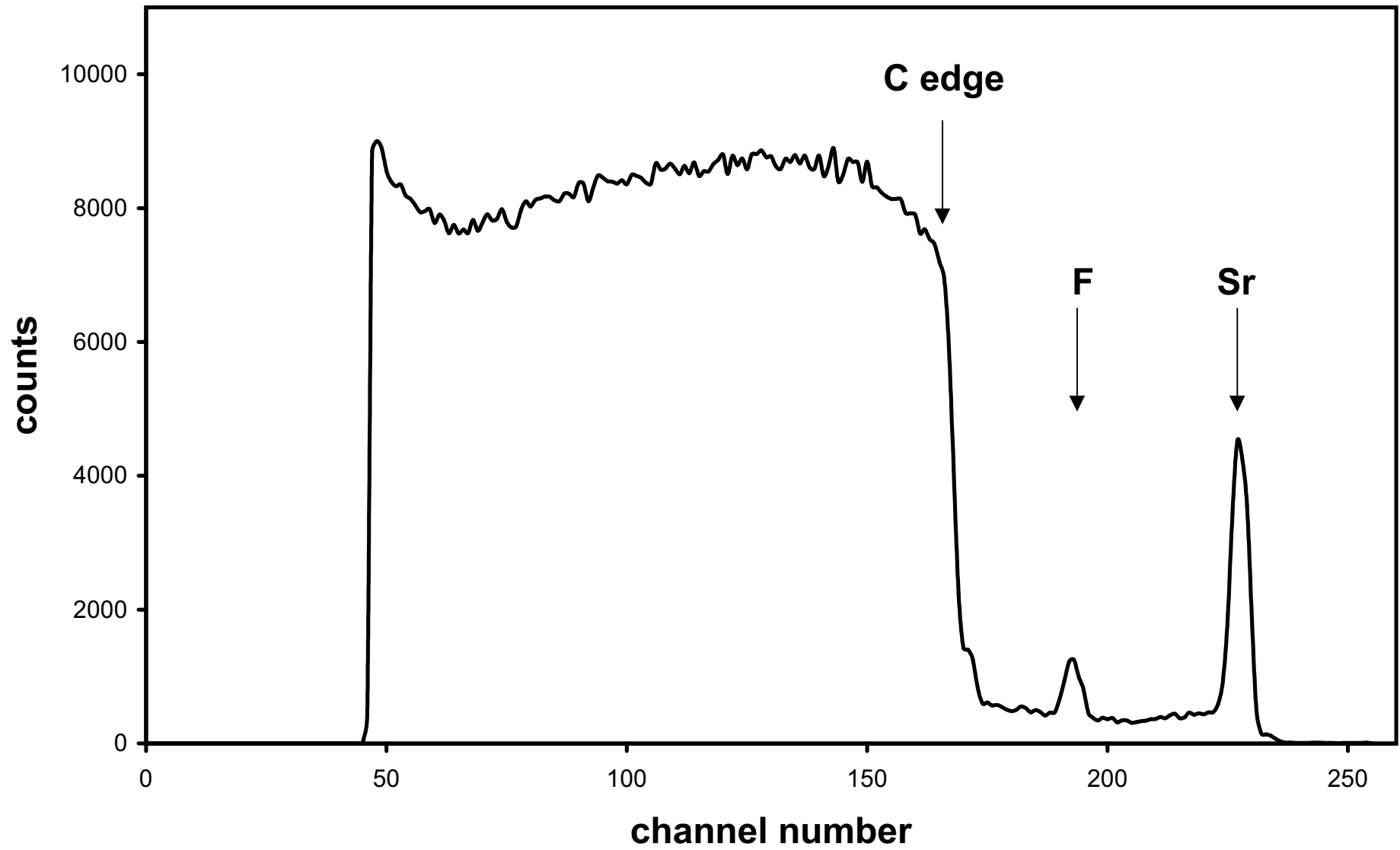
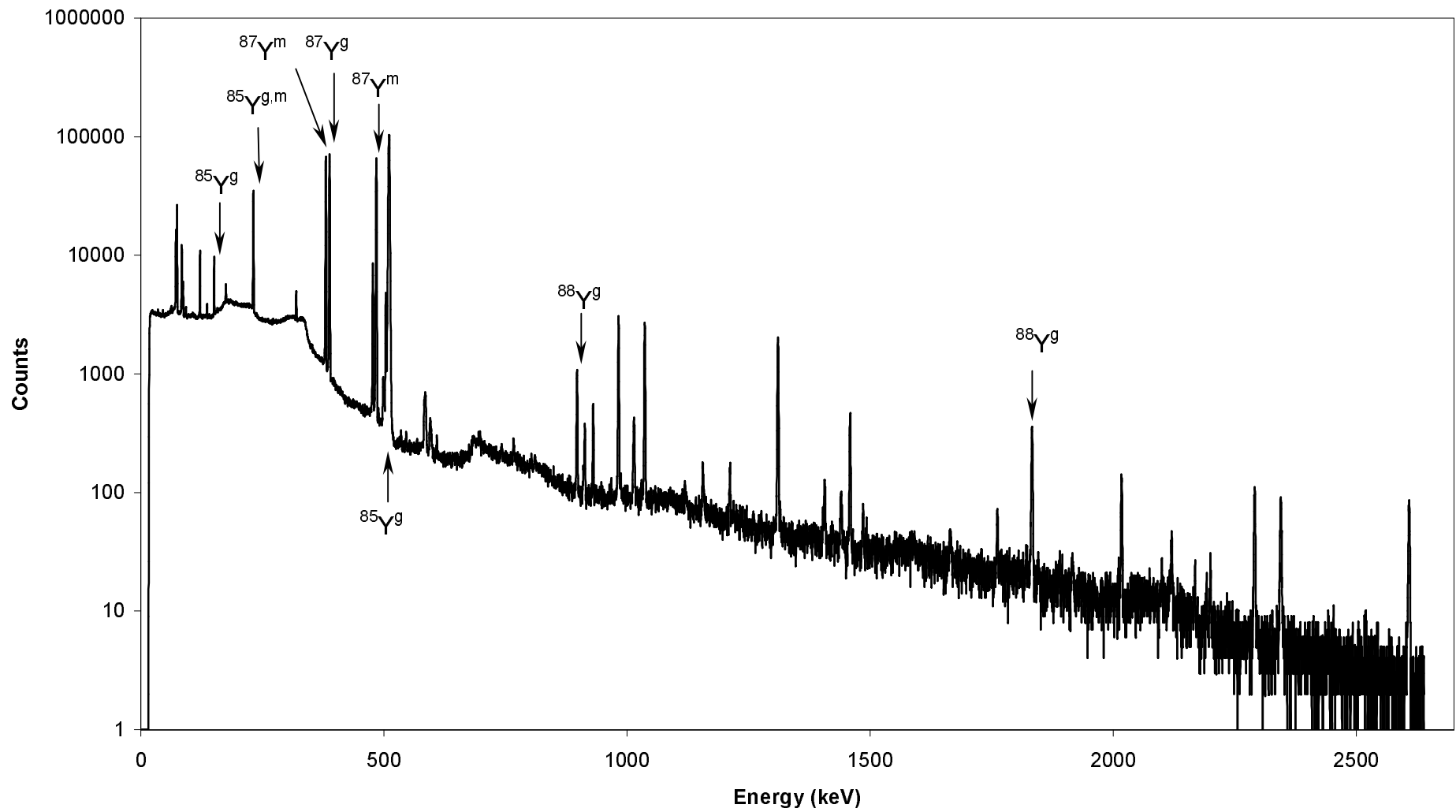
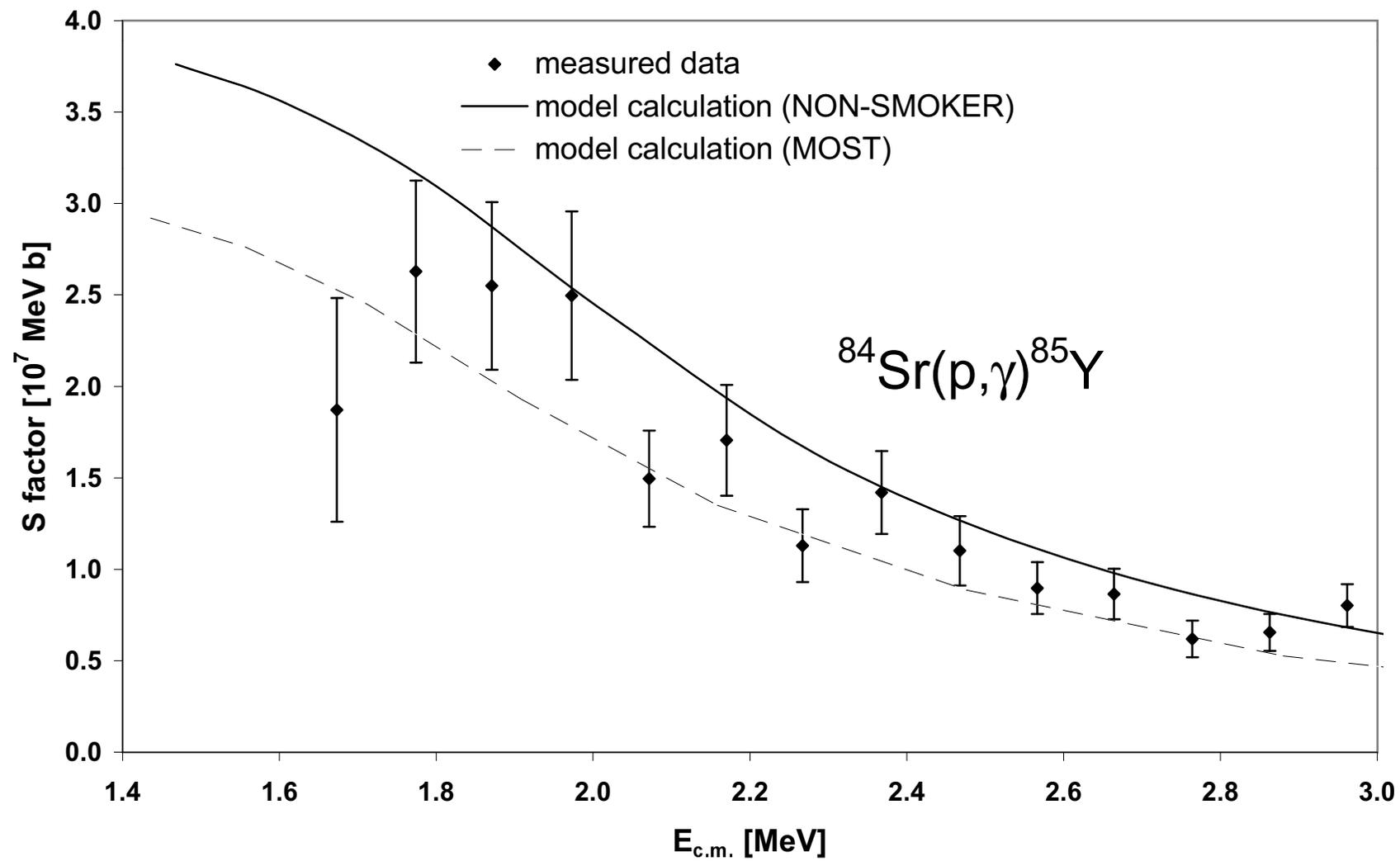
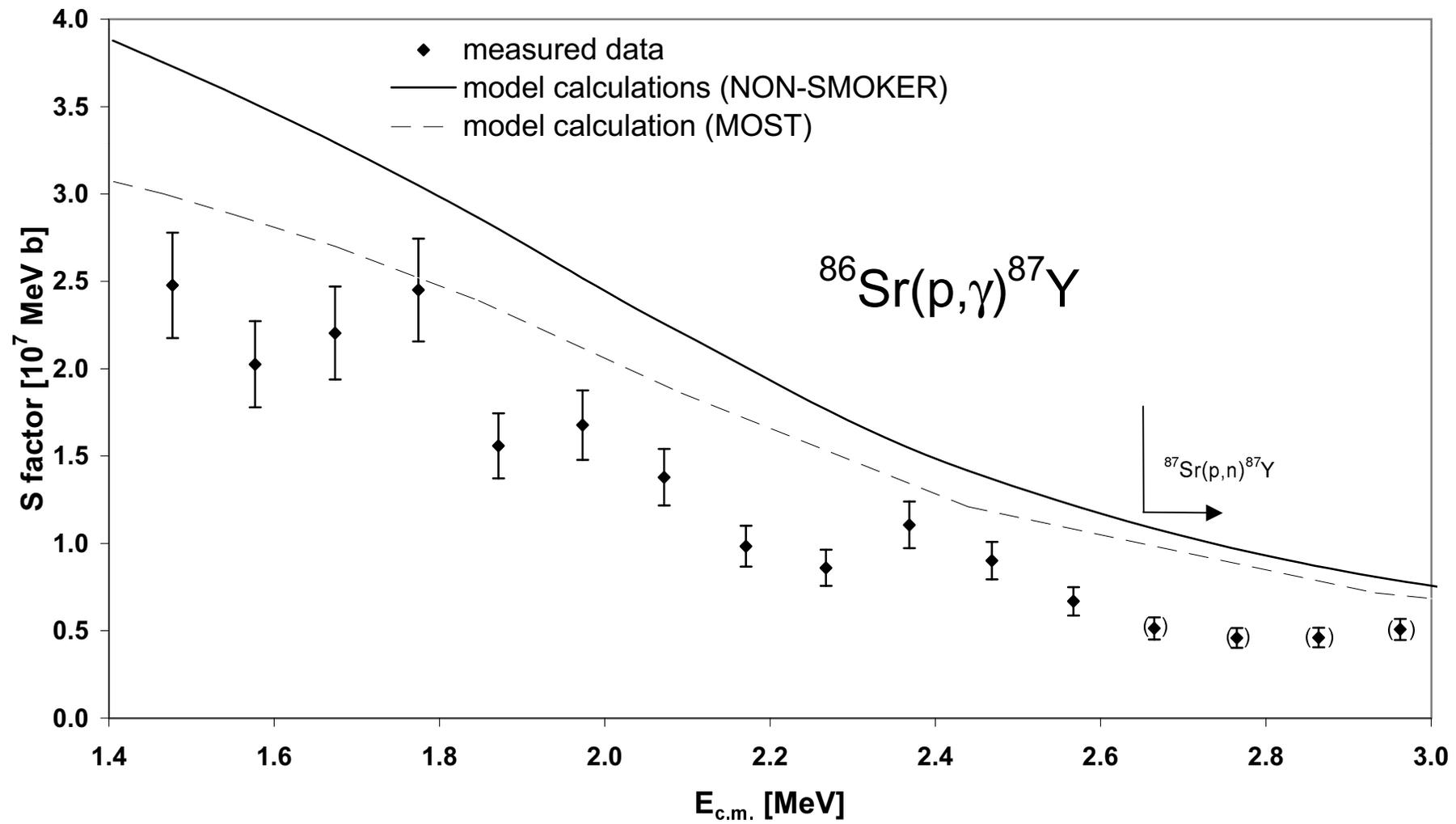
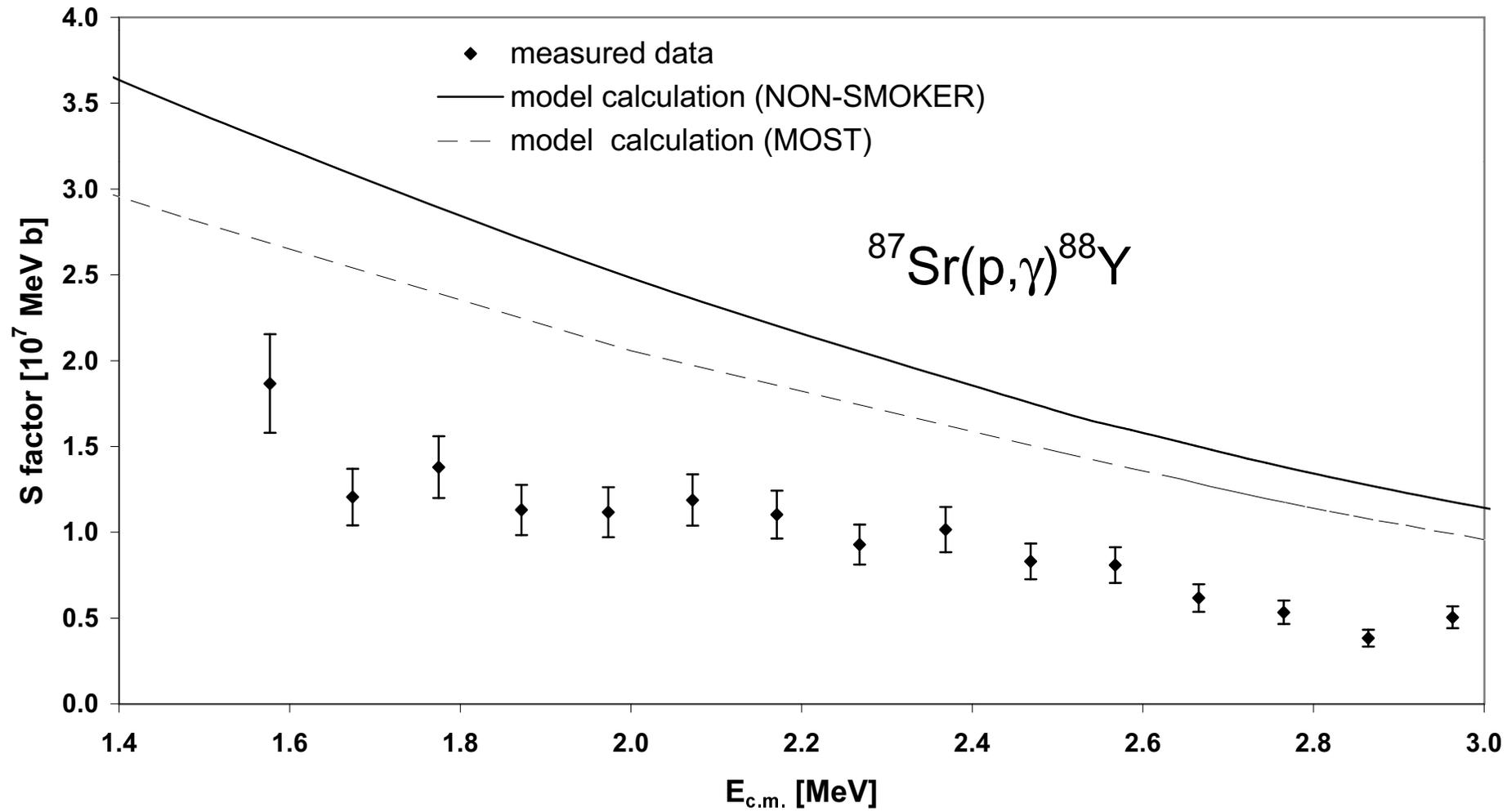
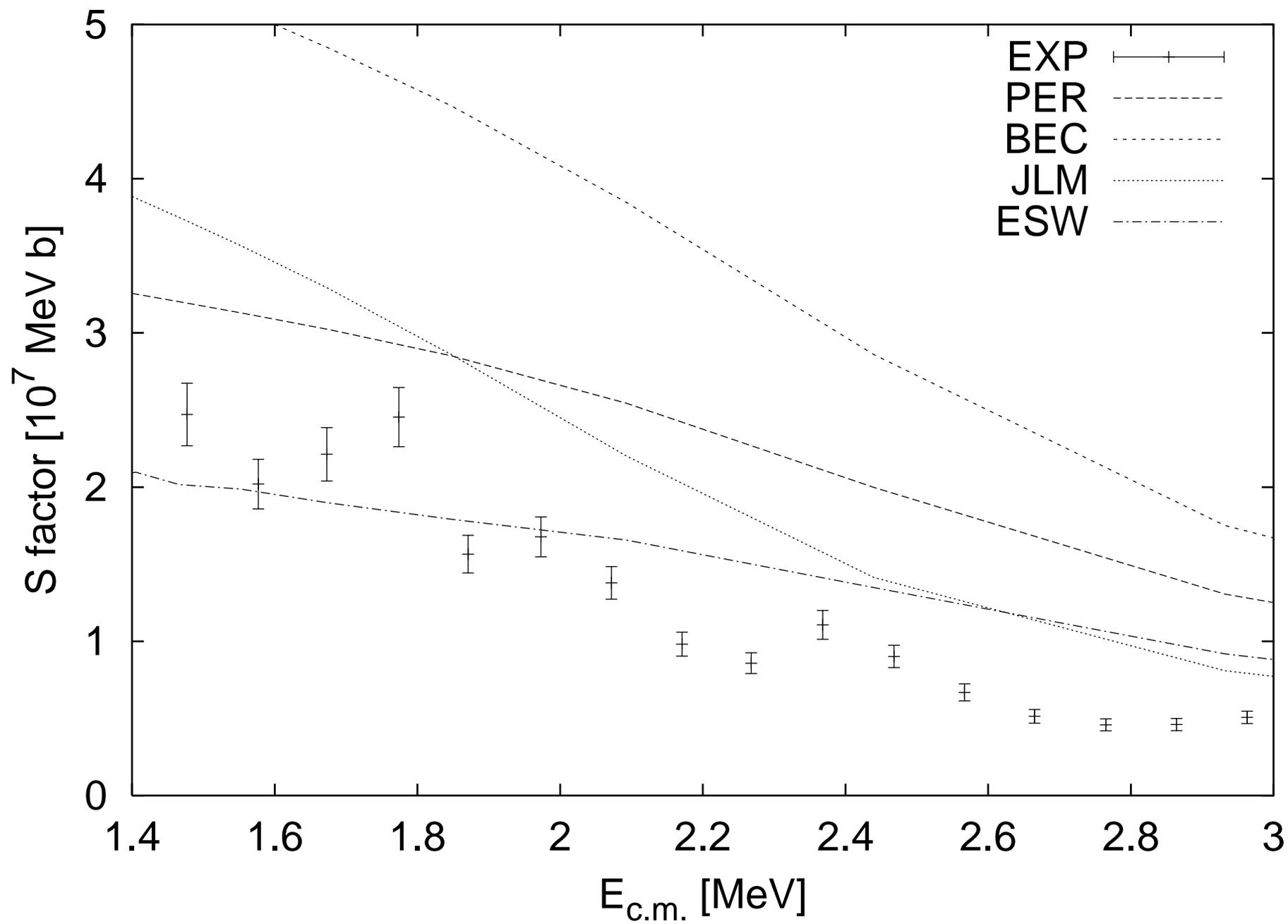
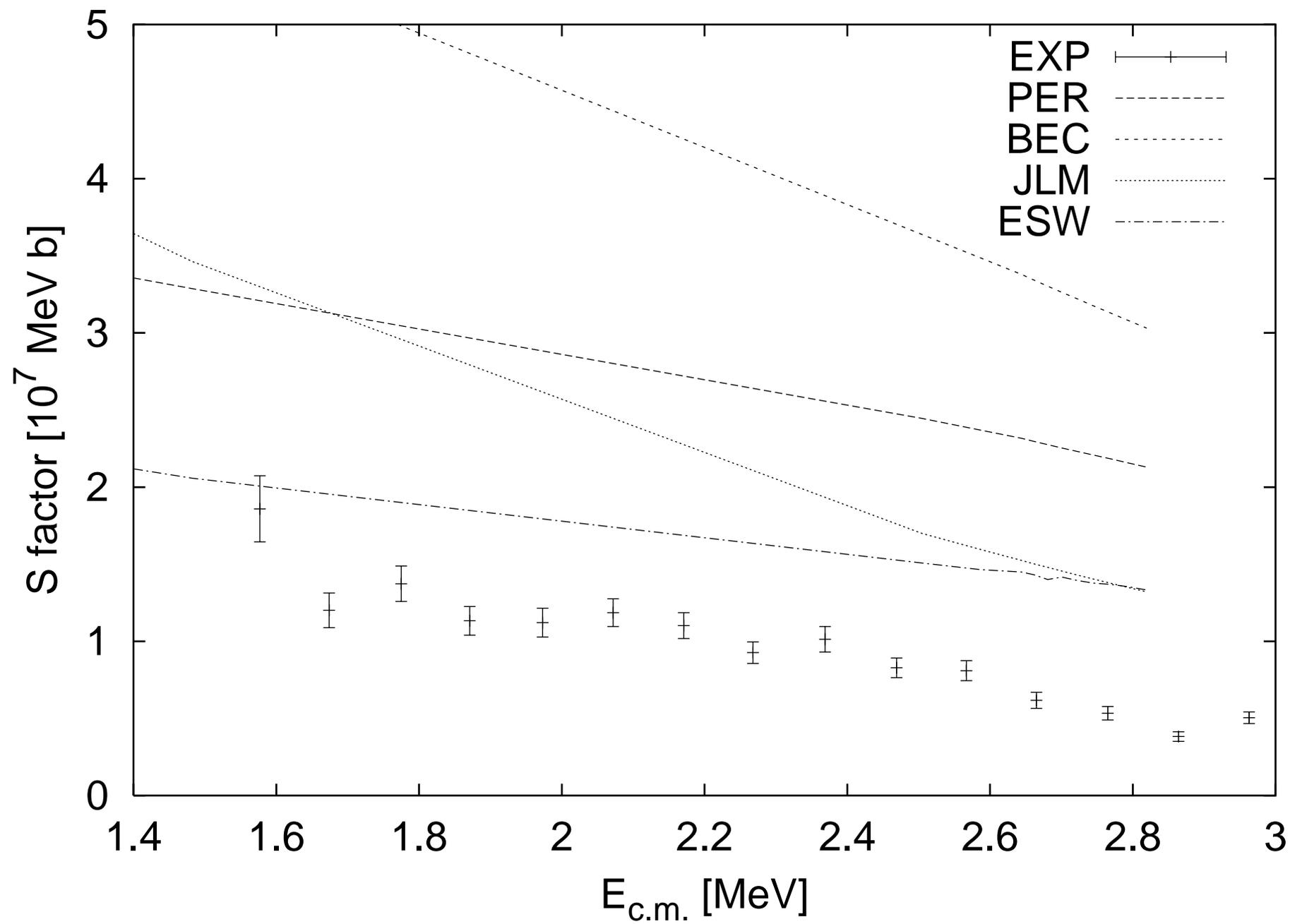
\end{document}